\documentclass[sigconf]{acmart}
\AtBeginDocument{%
  \providecommand\BibTeX{{%
    \normalfont B\kern-0.5em{\scshape i\kern-0.25em b}\kern-0.8em\TeX}}}
    
\usepackage{xac}

\setcopyright{rightsretained} 

\begin{document}

\copyrightyear{2021}
\acmYear{2021}
\acmConference[CHI '21]{CHI Conference on Human Factors in Computing Systems}{May 8--13, 2021}{Yokohama, Japan}
\acmBooktitle{CHI Conference on Human Factors in Computing Systems (CHI '21), May 8--13, 2021, Yokohama, Japan}\acmDOI{10.1145/3411764.3445547}
\acmISBN{978-1-4503-8096-6/21/05}

\title{Revamp: Enhancing Accessible Information Seeking Experience of Online Shopping for Blind or Low Vision Users}

\author{Ruolin Wang}
\affiliation{%
  \institution{UCLA HCI Research}
  %\city{Los Angeles}
  %\state{CA}
  }
%\email{violynne@ucla.edu}

\author{Zixuan Chen}
\affiliation{%
  \institution{UCLA HCI Research}
  %\city{Los Angeles}
  %\state{CA}
  }
%\email{zixuanchen@ucla.edu}

\author{Mingrui ``Ray'' Zhang}
\affiliation{%
  \institution{The Information School,\\ University of Washington}
  %\city{Seattle}
  %\state{WA}
  }
%\email{mingrui@uw.edu}

\author{Zhaoheng Li}
\affiliation{%
  \institution{Department of Computer Science and Technology, Tsinghua University}
  %\city{Beijing}
  %\state{China}
  }
%\email{lizhaoha17@mails.tsinghua.edu.cn}

\author{Zhixiu Liu}
\affiliation{%
  \institution{Computer Science Department,\\ Stanford University}
  %\city{Palo Alto}
  %\state{CA}
  }
%\email{liu16@stanford.edu}

\author{Zihan Dang}
\affiliation{%
  \institution{UCLA HCI Research}
  %\city{Los Angeles}
  %\state{CA}
  }
%\email{ritad@ucla.edu}

\author{Chun Yu}
\affiliation{%
  \institution{Department of Computer Science and Technology, Tsinghua University
  }
  %\city{Beijing}
  %\state{China}
  }
%\email{chunyu@tsinghua.edu.cn}

\author{Xiang ``Anthony'' Chen}
\affiliation{%
  \institution{UCLA HCI Research}
  %\city{Los Angeles}
  %\state{CA}
  }
%\email{xac@ucla.edu}

\renewcommand{\shortauthors}{Wang \textit{et al.}}


%% \renewcommand{\shortauthors}{Trovato and Tobin, et al.}

\begin{abstract}

  % \xac{the title says `blind or low vision' that is different from blind and visually impaired?}
  %\xac{note that the correct submission format is `\\documentclass[manuscript]{acmart}' but we can use whichever format we want before submission.}
Online shopping has become a valuable modern convenience, but blind or low vision (BLV) users still face significant challenges using it, because of: 1) inadequate image descriptions and 2) the inability to filter large amounts of information using screen readers. To address those challenges, we propose Revamp, a system that leverages customer reviews for interactive information retrieval. Revamp is a browser integration that supports review-based question-answering interactions on a reconstructed product page. From our interview, we identified four main aspects (color, logo, shape, and size) that are vital for BLV users to understand the visual appearance of a product. Based on the findings, we formulated syntactic rules to extract review snippets, which were used to generate image descriptions and responses to users’ queries. Evaluations with eight BLV users showed that Revamp 1) provided useful descriptive information for understanding product appearance and 2) helped the participants locate key information efficiently.

\end{abstract}






\begin{teaserfigure}
  \includegraphics[width=\textwidth]{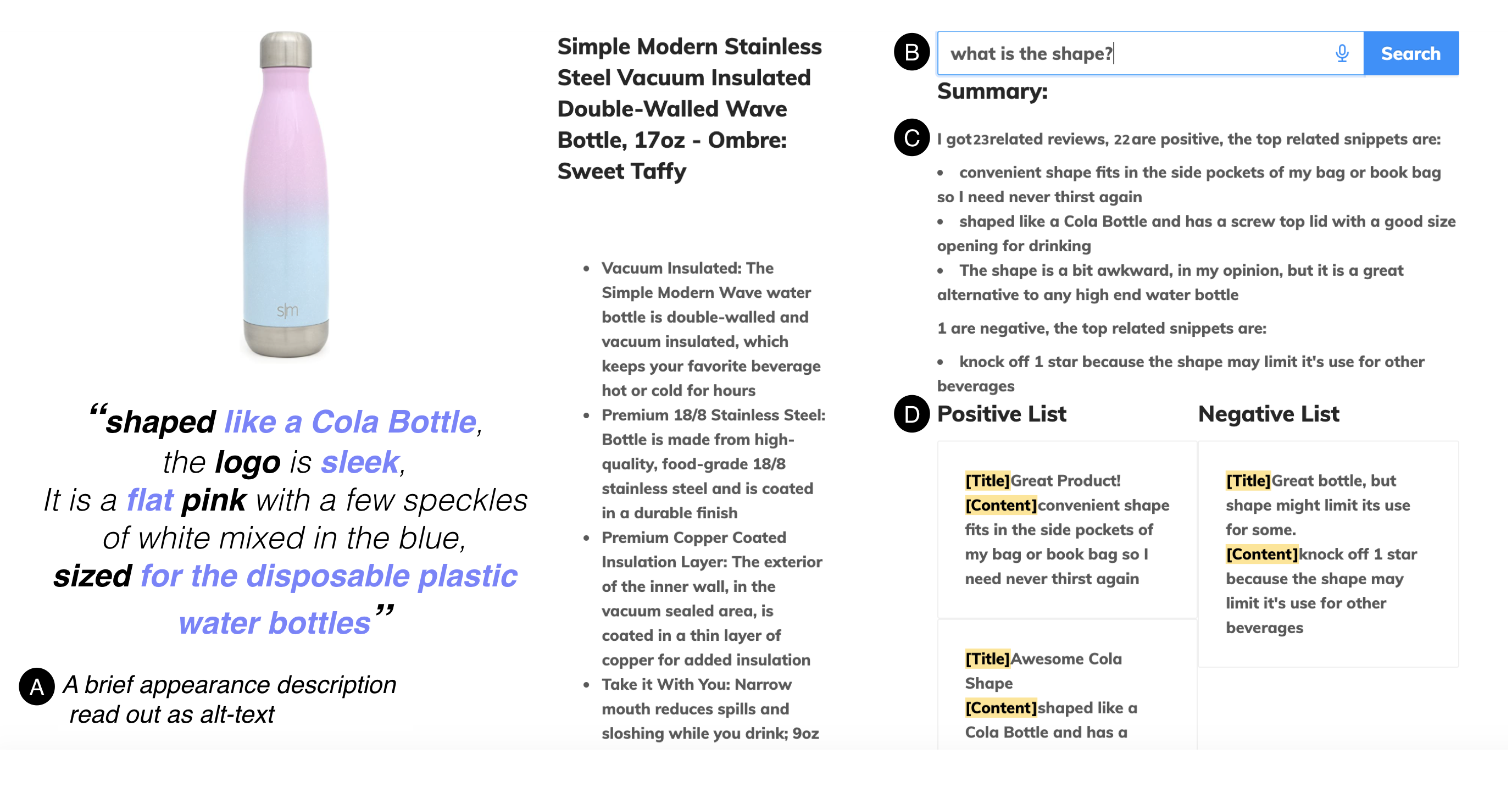}
  \caption{%Revamp is an interactive information retrieval system utilizing the customer reviews to help blind or low vision (BLV) users understand the visual information of a product. When triggered, 
  Revamp simplifies and reconstructs the original Amazon web page for Blind or Low Vision users' information seeking task to understand product details, especially the appearance. Using rule-based heuristics, Revamp extracts descriptive information from customer reviews to generate image descriptions (a), responses to users' queries (b) with a sentiment summary of all the relevant reviews (c) and original reviews sorted into a positive and a negative lists (d).
  }
  \label{fig:teaser}
  \Description{This figure shows a product web page modified by Revamp. In the left region, there is an image of this product, with a brief appearance description of it as ``'shaped like a Cola Bottle, the logo is sleek, it is a flat pink with a few speckles of white mixed in the blue, sized for the disposable plastic water bottles'; In the right region, there are the product name, basic information and review-based questioning answering function.}
\end{teaserfigure}

%figure1 -> rules

\maketitle

\section{INTRODUCTION}
% promise -------------------------------------------------
Online shopping has gained increasing popularity among Blind or Low Vision (BLV) people who have limited mobility to travel to physical stores. According to a research survey conducted in the UK, over 90\% of people with disabilities shopped online at least once a month \footnote{http://www.clickawaypound.com/index.html}.
The recent COVID-19 pandemic further accelerated the adoption of online shopping among BLV users \footnote{https://www.shropshirestar.com/news/health/coronavirus-covid19/2020/04/07/online-shopping-priority-plea-for-the-blind/}.
Thus making online shopping experience accessible has become an imperative requisite for ensuring the quality of life for BLV users.

% https://www.bbc.com/news/disability-52118942, https://www.blindveterans.org.uk/coronavirus-covid-19-updates/coronavirus-covid-19-assisting-a-blind-or-partially-sighted-person-with-shopping/,

% problem -------------------------------------------------
However, prior studies \cite{10.1145/3313831.3376404, Stangl:2018:BOC:3234695.3236337} show that BLV people still face significant barriers on online shopping websites due to inadequate image descriptions and screen readers' inability to filter out a large amount of information on a product page. 
% the passive reading approach of screen readers on massive information.
Our formative study with 20 BLV people showed that automatic tools (\eg Seeing AI, filters of screen reader) were only used by a few experienced users, and the information provided was too generic to inform a purchase decision, especially for certain categories where the appearance matters, \eg fashion products.
The most efficient way for BLV users is still seeking help from a sighted person, such as a family member or a crowdsourced helper, which is not always available.

% \xac{need to mention some closest related work, not just commercial products}

% proposed solution ---------------------------------------
To address the above challenges, we developed \textit{Revamp}, an interactive information retrieval system
that supports review-based question-answering (QA) on a reconstructed product page. It is implemented as a browser extension for Amazon.com as shown in Figure \ref{fig:teaser}.
% following three design implications derived from the interviews:
% \one{simplifying and reorganizing the webpage to better support a BLV user's task of understanding a product, especially its appearance;}
% \two{leveraging reviews containing detailed descriptions to fill in the inadequate visual information, and;}
% \three{responding to active queries with both a summary of and the original sources of reviews.}
% \xac{i feel the third point is the weakest and has been well practiced by industry. \eg google's zero-click response to search queries. how is this 3rd point specific to BLV users?}
% \xac{these three implications also seem a little repetitive compared to the descriptions below.}
%\xac{tentatively removed these parts to see how the intro would look}
Utilizing reviews as a knowledge source, Revamp extracts informative descriptions to help BLV users understand the product appearance.
To understand what kinds of visual questions BLV users care about, we collected questions from ten BLV participants on their frequently shopping categories on Amazon (Clothing, Shoes \& Jewelry, Home \& Kitchen, and Electronics), identified four main aspects (color, logo, shape, and size) in understanding the appearance. Based on the results, we formulated syntactic rules to extract the review snippets to generate image descriptions and responses to users’ queries.
%generating image descriptions (Figure~\ref{fig:teaser}a), responses to users' queries (Figure~\ref{fig:teaser}b) with a sentiment summary of all the relevant reviews (Figure~\ref{fig:teaser}c)
% and allows users to access the original reviews on and original reviews sorted into a positive and a negative lists (Figure~\ref{fig:teaser}d).

% proof ---------------------------------------------------
We tested the performance of these rules on the Amazon best-seller product lists covering three main shopping categories. Results showed that these rules \one covered 85\% informative reviews voted by BLV participants and \two provided insights on information conveyed by images evaluated by two sighted people.
We evaluated the usability of Revamp with eight BLV people on six representative products in the three aforementioned categories. Participants reported that Revamp reduces their efforts of information seeking, improves their utilization of reviews, and extracts reviews that are informative to help them understand product appearance. They considered Revamp to be helpful for shopping online independently when no sighted helpers were available.

Our {\bf contributions} are:
% contribution ---------------------------------------------------
\begin{itemize}
\item{Understanding the challenges faced by BLV people in information seeking when shopping online and design implications to meet their unique needs;}

\item{Identifying specific questions important to BLV users when shopping online and deriving syntactic rules that retrieve informative reviews to provide image description or to answer visual questions;} 

\item{Revamp, an interactive information retrieval system that integrates with Amazon to support product browsing and review-based question-answering;}

\item{Evaluation study with eight BLV users that validated the feasibility and usefulness of enhancing accessible online shopping experience via Revamp.} 
\end{itemize}
                                    
\section{Related Work}
This research builds upon prior work from different sub-disciplines across accessibility, information retrieval (IR), computer vision (CV) and natural language processing (NLP). 
We first review previous research investigating online shopping experience for BLV users, as well as the efforts to improve the accessibility of such experience; we then provide an overview of information retrieval methods that utilize online reviews to generate useful information, which serves as the foundation of the implementation of Revamp.
% \xac{logically, we usually start with general, then narrow down to a subset of users}

\subsection{Online Shopping Accessibility}
Blind or low vision (BLV) people
% 's strong desire of being treated as ordinary made them 
often take great effort to find and learn what products are visually appropriate for them \cite{Liu:2019:BTM:3290605.3300602, Shinohara:2011:SMA:1978942.1979044}. 
Without real-time communication with store clerks and touching the products directly, information related to online products is mainly conveyed through images and text provided by sellers. Hence, inadequate descriptions of images and unparsed text information significantly reduce BLV consumers' engagement with online shopping websites \cite{10.1145/3313831.3376404, Stangl:2018:BOC:3234695.3236337}.
Currently, BLV people could leverage human-powered assistive tools to answer visual questions about daily objects and products in stores \cite{Bigham:2010:VNR:1866029.1866080, Avila:2016:RAB:2910674.2935839, Brady:2013:VCE:2470654.2481291, Morris:2014:RSA:2531602.2531707, Yuan:2017:IDK:3171581.3134753, Rodrigues:2017:IQS:3132525.3132555, 10.1145/2384916.2384941}. Specially in shopping clothes, human helpers can provide remote assistant on describing articles of clothing (\eg color) \cite{Avila:2016:RAB:2910674.2935839, Brady:2013:VCE:2470654.2481291} or offering subjective fashion advice \cite{Morris:2014:RSA:2531602.2531707, 10.1145/2384916.2384941}. 
Relatedly, automatic photo caption generation and visual question answering in computer vision and AI research is often designed for general scenarios \cite{10.5555/1888089.1888092, ramnath2014autocaption, jiang2018pythia}. There exist automatic solutions that promote accessible image understanding on the web, 
\eg providing image alt-text for photos in Facebook based on object recognition \cite{wu2017automatic}, 
using OCR for photos on Twitter and other websites alike \cite{10.1145/3308561.3354629, 10.1145/3313831.3376728}, 
and using reverse image search to find existing captions \cite{guinness2018caption}. 

% crowdsourcing mechanism for evaluations \cite{Lasecki:2014:LSR:2661334.2661352, chen2015microsoft, jas2015image, salisbury2017toward, von2006improving}. (but not specially for online product images

%It is unknown how imperfections in emerging tools for automatic caption generation may help or hinder BLV people's understanding of online product appearances.

However, such solutions cannot fulfill BLV users' special need in better understanding visual details of online products.
Most images on eCommerce websites, if captioned by emerging automatic caption tools such as SeeingAI \footnote[1]{https://www.microsoft.com/en-us/ai/seeing-ai}, will only have a generic description of the product, \eg “\textit{probably a close up of a red chair}”, without any specific insights on appearance details, \eg how is the size, and how does this kind of red make people feel.
BrowseWithMe \cite{Stangl:2018:BOC:3234695.3236337} has made an important step forward in describing clothes outfit by identifying the image regions, but only respond with basic color names (\eg, ``\textit{Brick Top, Navy Pants}'') based on color detection of image. 
Our contribution is providing a novel perspective of leveraging reviews as an external information source to supplement the visual details. 
%\xac{it'd be stronger to end with a statement that says how our work goes beyond these solutions}

% limitation: only contain one product on a clean, solid-colored background, and the image description desired by BLV people should include descriptive details of the product and varied based on the type of products\cite{10.1145/3313831.3376404}. 

% Accessible Information Seeking has been a rising topic
Besides inadequate descriptions of images, the online shopping experience of BLV people is also hindered by the weaknesses of screen readers dealing with crowded websites. 
Screen readers provide fine-grained information navigation and control, but at the cost of reduced walk-up-and-use convenience \cite{10.1145/3308561.3353773}.
Prior work augmented the auditory web browsing experience by adding a secondary voice output \cite{Sato:2011:SAV:1978942.1979353, Zhu:2010:SAV:1878803.1878870}, supporting basic queries about the information provided by sellers \cite{Stangl:2018:BOC:3234695.3236337}, and supporting users to choose how many and which levels of detail to listen to based on their interest \cite{10.1145/3173574.3173633}.
voice assistants as an alternative solution lack the ability to deeply engage with content and to get a quick overview of the landscape (e.g., list alternative search results \& suggestions) \cite{10.1145/3308561.3353773}.
Current voice assistants for online shopping such as \textit{Alexa} only respond to a limited range of queries, \eg “\textit{add bananas to the chart}”, rather than finding targeted information to answer a BLV user's question.
Inspired by VERSE \cite{10.1145/3308561.3353773} which extends a voice assistant with screen-reader-inspired capabilities to enhance web search, our work integrates review-driven question-answering with two levels of details (summary \& original review lists), which contains rich first-hand insights from other buyers.

\subsection{Information Retrieval In Online Reviews}
Most image captioning and visual question answering benchmarks to date focus on questions such as simple counting  and object detection that are directly based on the images. Marino \etal \cite{marino2019ok} ~ draws on external knowledge resources when the image content is insufficient to answer customers' questions, but their focus on  images containing general scenes cannot be directly leveraged for online shopping. 
Online reviews have been mentioned as an useful resource to assist BLV customers' desire for more product information \cite{Stangl:2018:BOC:3234695.3236337, Liu:2019:BTM:3290605.3300602, 10.1145/3313831.3376404}, such as supporting question-answering\cite{mcauley2016addressing, wan2016modeling} and generating summaries \cite{li2010structure, Huang:2012:REI:2380116.2380120}.
Prior work towards general users investigated extracting experiences \cite{min2012identifying, nguyen2012detecting}, tips \cite{guy2017extracting, zhu2018unsupervised} or snippets suitable for product descriptions \cite{10.1145/3308558.3313532, 10.1145/3357384.3357984}, yet it remains unknown how reviews can help with the unique interests of BLV users, especially providing descriptive information on product appearance. Our research fills a gap in the literature by leveraging existing human-authored resources of online reviews as an additional source of information to address such unique needs.

The most related work \cite{10.1145/3308558.3313532} applies a supervised method to extract reviews for image description based on 25K labeled sentences.
However, the descriptions generated by this work are generic and rarely covered descriptions of visual details. Rather than following supervised methods such as conducting aspect-based sentiment analysis utilizing feature engineering \cite{wagner2014dcu, kiritchenko2014nrc} or Bidirectional Encoder Representations from Transformers (BERT) \cite{sun-etal-2019-utilizing} that requires a huge amount of human-labelled data, we proposed empirically-formulated syntactic rules based on our studies with BLV participants to retrieve meaningful review snippets for understanding appearances.
Compared to ``black box'' data-driven approaches, these syntactic rules are explainable and can be validated by our targeted users; they are modular---existing rules can be edited or removed and new rules can be added without affecting the others. To the best of our knowledge, there is no prior work that directly informs rule design on extracting visual descriptions from reviews.

In summary, Revamp extends prior work on enhancing accessible information seeking by proposing a solution to bridge customer reviews with the needs of BLV users on visual information for a wider range of online products, including three frequently shopped categories: Home \& Kitchen, Clothing, Shoes \& Jewelry, and Electronics.

\section{Overview of Studies}
In this work, we conducted three groups of studies:

\begin{itemize}
    \item{\textbf{Formative study (Section 4, 5)} with two stages to: (i) understand the current practice and derive design implications via 30-mins semi-structured interviews with 20 participants (P1-P20); (ii) investigate what specific visual information BLV users expect and how reviews can help via questionnaires and 30-mins semi-structured interviews with 10 participants (P11-P20).}

    \item{\textbf{Rule evaluation (Section 7.1)} with two stages to: (i) evaluate the informativeness of reviews extracted by the rules via questionnaires with eight participants (P13-P20); (ii) evaluate the rules’ generalizability on different products where two sighted researchers rated the quality of the generated alt-text and answers in three informative levels.}

    \item{\textbf{System evaluation (Section 7.2)} to evaluate how the Revamp prototype affected the shopping process. Eight participants (P13-P20) browsed products with Amazon web pages and Revamp then provided subjective comments by online interviews. The study lasted for 40 mins.}
\end{itemize}

The time interval between each group of study was about one month. Twenty participants (Gender: 7 female and 13 male; Age: average = 27.7, SD = 5.5) came from two different cultures (P1-P10: China; P11-P20: North America) participated in the first-stage formative study due to the distributed nature of the research team. Although the platforms and languages they use differ, participants share similar screen reader experience and information seeking challenges since the screen readers follow the general design and most online shopping websites are similarly structured with overloaded information. Only North American users were involved at the following studies because of the limited availability of review data on the Chinese shopping sites. Participants were compensated with \$15 USD\textbackslash hour. Each study was audio recorded, transcribed and coded by three of the authors following the reflexive thematic analysis methods from Braun and Clarke \cite{doi:10.1080/2159676X.2019.1628806}.
The demographic information of participants is attached as Appendix.

%The limitation of only involving eight participants in evaluation can be addressed by a large-scale in-the-wild study, which we will discuss as future work.

\section{Understanding Current Practice and Deriving Design Implications}
We conducted 30-minute semi-structured online interviews with 20 BLV consumers (P1-P20) recruited through social media platforms. Specifically, the goal of this study is to understand that, when shopping online:
\begin{itemize}
  \item What challenges do BLV users face in information seeking and what are their coping strategies?
  \item What are the design heuristics if we want to build an accessible information retrieval system for BLV users?
\end{itemize}

\subsection{Challenges}
Aligned with prior work \cite{10.1145/3313831.3376404, Stangl:2018:BOC:3234695.3236337}, two main challenges mentioned by our participants are lack of visual information and lack of efficient ways to navigate through information.
We revealed how the two challenges were intertwined with each other, which elicited insights on designing an efficient way to leverage reviews to fill in the lacking visual information.
%related with BLV users' specific needs on understanding product appearance under the context of online shopping.

\subsubsection{Lack of Visual Information}
The most frequently mentioned challenge is the lack of detailed information for understanding the visual appearance of the product (18 out of 20 participants). While sighted users can tell the visual attributes such as color, shape and size or even functionalities of a product from a single image, 
%BLV users depend on the alt-text of the image to understand its content. However, 
the alt-texts of many current product images are either empty, or set as the image path, which does not provide any visual information for BLV users at all.
Sellers usually provide no textual descriptions equivalent to the visual information conveyed by images, whether it is some basic attribute such as color and shape, or some vivid details such as pattern design.
Take the color attribute as an example: the names provided by sellers sometimes can be too generic or obscure to interpret accurately.
So much of the language we use to describe a product is centered around visual, \eg \code{marble pattern} or \code{Arctic blue}, which poses a barrier to people who don’t experience the world visually.
% \xac{up to this point, the writing seems to come from the authors' experience as there is no mentioning of participants}
%P5 mentioned his story that once he wanted to buy a guitar and was curious about what exactly was the color attribute `gold'. \code{Is it a shining one or a matte one? I couldn't figure it out by myself unless there had (been) more descriptions.}
P18 brought up an example that a T-shirt with the color name \code{surf the web} was actually a kind of \code{bright blue}, which was only mentioned by a review.

\subsubsection{Lack of efficient ways to navigate through information}
Comparing with the scanning and skimming experience of sighted users, the passive and sequential reading manner of screen readers makes it inefficient for BLV users to retrieve useful information scattering all over a web page.
P16 mentioned that it can be frustrating when accidentally jumping into another unrelated web component such as the shopping cart in the process of exploring product details. 
%which is overloaded with reviews, product information and details provided by sellers, etc. 
%\code{I know there exists a lot of useful information on the product page, but it is too clunky to find them in the clutter.} (P14)
All participants agreed that reviews were useful resources. Review snippets with detailed descriptions can also provide further insights for product appearance.
%However, sifting through the large amount and often duplicated reviews can be especially laborious. 
%hence the detailed descriptions and subjective feedback from other buyers are under-utilized, including those that can provide further insights for product appearance.
%\eg “\textit{the light-colored marble pattern feels cool to use in summer}” and “\textit{this elegant marble pattern suits the white wall}” (P3).
However, most participants (16 out of 20) reported their utilization of reviews was greatly reduced because sifting through the large amount and unrelated reviews can be especially laborious. P19 stated that ``\textit{Usually there are too few reviews per page and there are no easy ways to jump from review to review. I know there exist useful reviews but it's hard to directly pull them out without efforts.}''. 

\subsection{Current Solutions}
There still exists a gap between current solutions and BLV users' expectations of online shopping experience. Relying on human helpers, the same as many other scenarios, suffer from the limitation of availability, raises privacy concerns, and reduces BLV users' independence. 
Meanwhile, as reviewed earlier in Related Work, many existing techniques lack adaptation to the specific needs of BLV users' online shopping.

\subsubsection{Seeking Help from Sighted People}
80\% of the participants chose to seek help from sighted people including families/friends, other customers, paid video chat and accessibility hotlines.
In this way, they had to trust the helper's personal judgment and subject to issues such as helpers' inavaliability and privacy concerns.
P19 mentioned that \code{The shipping and delivery process is not the most time-consuming for me. Waiting for someone to help me with selecting products is.}
P12 stated that he did not always want others to know about certain products he buys and hoped there was an alternative solution to shop more independently.

\subsubsection{Leveraging Existing Web Functions and Automatic Tools}
P18 mentioned using visual interpretation apps such as Seeing AI to get brief generic descriptions on product images based on object and color detection, \eg \code{probably a close up of a red chair}.
%The only one participant with low vision mentioned that she usually uses image processing tools to check the RGB values of the color. 
Two participants who were more tech-savvy in web interfaces (P11, P20) mentioned using keyword-based search and filters, but these tools are under-utilized by other participants and can only remove irrelevant information on a general basis.
P8 and P13 mentioned the customer questions \& answers section supported by websites can also help with some generic details, yet it seldom covers appearance-related information, which is assumed available to sighted consumers via product images.
%  can capture a product's visual appearance directly from images.

%The customer questions \& answers section supported by websites hardly cover appearance-related discussions since sighted consumers can capture a product's visual appearance directly from images;
% Searching by keywords and using filters could help remove unrelated information, but cannot pick out the reviews containing detailed descriptions which can provide further insights to appearance.

\subsection{Implications for Design }
%\xac{suggest: Implications for Design. a heuristic would be "Product page should make it easy for BLV people to navigate information"; an implication could be a heuristic and/or a solution to realize a heuristic; what's described below sounds more like solutions}

Based on the interview results, we derived 3 design implications to improve the accessibility of online shopping experience. Specifically, the designs focus on leveraging the existing resources on a product page: 
\textit{simplifying the webpage} (4.3.1) and \textit{responding to active queries} (4.3.3) are to address the \textit{lack of efficient ways to navigate through information} (4.1.2); \textit{leveraging reviews to supplement visual description} (4.3.2) is to address the \textit{lack of visual information} (4.1.1).

\subsubsection{Simplifying and reorganizing the webpage structure}
When the users’ current task is understanding a product, especially the appearance, the overloaded information and web components can become burdens for seeking visual-related description. To address this problem, we can provide an alternate view of the original page with less related information and components removed, such as shopping cart. Meanwhile, it is important to provide users the option to switch back to the original page.

\subsubsection{Leveraging customer reviews for extra visual description}
%\xac{might need to iterate more ...}
%All of our participants agreed that the customer reviews was a helpful resource for understanding the product, yet they also found it was extremely tedious to use. It is thus important to extract the information and present it in a way that is easy for BLV users to navigate.
%leveraging other sources on the webpage to fill in the information asymmetry; Leverage Existing Reviews: 
Participants' responses suggest that some subjective information in reviews can be helpful for BLV people to understand the visual attributes. Just as P2 stated: "\textit{It is impossible for me to build a visual concept as you sighted people do, but I can feel what you feel. Reviews talking about the feelings when seeing the appearance are helpful for me to understand this pattern.}" It remains to be investigated how to retrieve the informative reviews to meet the specific needs of BLV users. 

\subsubsection{Responding to active queries with the page resources}
% should discuss the summary and sentiment
% codes from previous paper
%Instead of passively browsing the information on the product page, 
Participants expressed preferences to actively asking questions and getting summative answers of a product, which was usually what they did when consulting with a sighted helper.
%Participants mentioned that responding to their specific questions is more natural way of gathering information required for making a shopping decision.
%``\textit{I can understand how the search bar works but not everyone is good at asking good questions or selecting suitable keywords to search.}'' (P17) 
%Specific questions on appearance should be supported. %``\textit{The other consumers don't ask a lot about what I am concerned, for example, the appearance. To get satisfactory answers, I have to post a new question mentioning that I am blind and please describe the details for me. Also, I am not sure how long I will wait until someone answers.}'' (P8) 
%``\textit{Some reviews are understandably visual in nature and don’t answer my questions about what exactly the product looks like to my liking.}'' (P13) 
Meanwhile, any additional assistance mechanisms should be compatible with users' current screen reader experience to support users to access the original information as well. P6 commented \code{I do not completely trust in answers selected by machines in case it may not cover very well}.
%Only providing reviews snippets may not be informative enough, ``\textit{The context of reviews can be very helpful, and I often get unexpected details about other aspects of the details.}'' (P1)
%\cite{10.1145/3308561.3353773}

\section{Leveraging Reviews to Answer BLV Customers' Questions}
%respond to queries
% It remains unknown how 
The first-stage formative study uncovered the opportunities and challenges to support the unique needs of BLV population in visual information seeking in online shopping.
%supplement a lack of visual information by mining information from the customer reviews.
%To support the unique needs of BLV population in information seeking
% for online products of broader categories by leveraging the existing information on shopping websites, 
Building upon these findings, we then seek to further understand:

\begin{itemize}
  \item What specific visual or other product-related questions are BLV users interested in?
  \item How can we retrieve useful review snippets to answer these questions?
\end{itemize}

% \xac{do we need to mention that reviews are also used to generate image description? which doesn't have a question to start with and need to be treated slightly differently}
% \vio{is it ok to just mention it later in 4.4? after we found that users started with general appearances questions then detailed questions, we decided to support both answering and a brief image description}

%In this work, we targeted at North America blind consumers (P11 - P20) who mainly use Amazon for online shopping. 
It is beyond the scope of this paper to exhaustively cover the large number of product categories and all the consumer questions corresponding to each category.
To narrow down our focus, we first distributed questionnaires to ten North American participants (P11 - P20) who mainly use Amazon to collect their most frequently shopped product categories, then conducted one-on-one interviews to formulate the understanding of their specific questions on three main categories: Home \& Kitchen, Clothing, Shoes \& Jewelry, and Electronics, based on which we derived rule-based solutions to extract informative review snippets.

\subsection{Product Categories and the Corresponding Question Types}
The frequently shopped categories on Amazon by our participants are: Electronics (chosen by 90\% participants), Home \& Kitchen (50\%), Pet Supplies (50\%),  Audible Books and CDs (50\%), Clothing, Shoes \& Jewelry (30\%), Grocery and Gourmet Food (30\%). Among these categories, we focused on three main categories whose appearances are as crucial as other information for making purchase decisions,
including Home \& Kitchen, Clothing, Shoes \& Jewelry, and Electronics. 

We then gathered 168 questions from participants in these three categories.
Each participant was shown representative products on the Amazon best-sellers list and was asked to raise as many questions they want to ask as possible after browsing the title and description provided by sellers.
The questions were then labeled and divided into two groups,
non-visual questions and visual questions. The distinct breakout of visual (57\%) and non-visual (43\%) questions is based on whether a question can be answered by directly observing the image.

First, 72 out of the 168 questions were questions on non-visual information that shares common interests with sighted consumers, including
\one functionality and other non-visual attributes, \eg material, price; and
\two consumer feedback, \eg whether the product was worth the price, or of high quality.

We are interested in the remaining 96 questions on visual information, which were of specific interests to BLV users, including 
\one visual attributes of image, \eg color, logo, shape, size; and 
\two high-level concepts that can be inferred from an image, \eg usage method, style. 

\subsection{Types of Visual Questions and How to Find Informative Answers}
For non-visual questions, all participants reported that the useful information usually could be found in the product description provided by the sellers, reviews or customer questions \& answers, while for visual questions, they mainly rely on human helpers to get the answers.
Thus we focus our investigation on the following sub-categories of visual questions and whether we can provide informative answers to these visual questions based on reviews, hence providing an alternative for BLV users to shop online independently.

\subsubsection{Visual questions on basic attributes}
Most participants started with general appearance questions, such as \code{how does it look like?}, followed by detailed questions around the basic visual attributes of the product, including color, shape, size and logo, which puts forward the need of providing a briefing of appearance covering these basic attributes first, then responding to specific queries.
Participants mentioned that they are not satisfied with only knowing the basic category name of visual attributes, or just vague comments, \eg \code{Nice shape. This color looks great}. They feel interested in the specific detailed descriptions and impressions on appearance.
P5 mentioned his story that once he wanted to buy a guitar and was curious about what exactly was the color attribute `gold'. \code{Is it a shining one or a matte one? I want to know more details.}
For shape and size, participants always want to know how to compare the product with daily objects, \eg \code{Does it fit on a tabletop?}, \code{Is it like the size of a banana?}, which can help them better understand shape and size in the real world.
Participants are also curious about if there are logos on the product.

To provide informative answers on basic attributes mentioned above, the key is to extract the descriptive and comparative details on appearance from the large amount of reviews. Particularly, clothes can have more nuanced design details to describe, such as sleeves, waist, neck, edition of clothing, graphic designs or patterns. These questions, \eg \code{Do the buttons go all the way down? What kind of neck design does the shirt have?}, are very category-specific and often require fashion knowledge to provide aesthetic descriptions, as P14 mentioned \code{even my friends don't describe well}, hence descriptions in common customer reviews hardly meet their expectations. 
In this paper, we don't focus on these subdivided fashion knowledge which requires large amount of annotated data (\textit{e.g.} DeepFashion \cite{liu2016deepfashion}), but focus on the common attributes shared by a wider range of products and possibly covered by customer reviews. The four attributes (color, shape, size, logo) are common across many products and can be concisely described in the alt-text compared to other aesthetic attributes (e.g., pattern, style).

% which is beyond the scope of this work. 

%\begin{itemize}
%    \item fit with other clothes, or fit with the specific occasions (speaker in room, bottle for hiking, shoes for business)
%    \item expectation for a certain function (yes or no question), and function details (technical details)
%    \item if many reviews why so popular
%    \item comfortable (general)
%    \item quality and durability (general), product care
%    \item material
%    \item size dimensions (shoes, clothes length)
%    \item date release
%    \item electronics devices: lightening
%\end{itemize}
%

\subsubsection{Visual questions on high-level concepts}
There are also high-level questions not directly related to basic visual attributes but can be inferred from the appearances in the image with commonsense, including usage, style, quality, texture and specific accessibility requirements, \eg, \code{Does it look like it has good quality? How to open this bottle? Does the product have physical buttons?}
Such properties are similar to the ``signifier'' concept proposed by Don Norman
\footnote{https://sites.google.com/site/thedesignofeverydaythings/home/signifiers}, 
which act as the indicator that can be interpreted meaningfully in the context of the social and physical world.
These questions, usually framed as Visual Commonsense Reasoning, has been a challenging research topic in computer vision \cite{zellers2019recognition}. Yet currently it is not well-explored to address the need of inferring concepts and functions from product images for online shopping scenarios, as they require a large amount of annotated data from external knowledge sources beyond the shopping websites. However, the reviews commenting on the specific aspects (\textit{e.g.} quality, physical buttons) can still indirectly provide insights on customers' opinions. Thus our expediency is treating these questions same as the non-visual questions to provide the relevant reviews.

\subsection{Rules-Based Solution for Retrieving Informative Review Snippets}
In this work, we mainly focused on addressing BLV users' questions on the common basic visual attributes, including color, logo, shape and size, which can be supported by augmenting keywords-based review searching with rule-based filtering (described below) to extract more informative snippets.

Compared to data-driven models, rule-based solutions can work well without the need of manually labelled data. Further, rules are explainable and can be validated by our targeted users. Rules are also modular---existing rules can be edited or removed and new rules can be added without affecting the others.
Note that currently our rules are not designed to answer questions related to high-level concepts, which we consider as future work.

% we currently filter out unrelated reviews using keywords.
%Sentiment: Some review sentences that express only sentiment comments on the product are not informative enough for better understanding the visual attributes. Examples include ``\textit{This color is great}'', ``\textit{The product is in good shape}''.
% and support basic keywords-filtering for other categories of questions

%Since we don't have enough data for data-driven methods, we adopted the following method: We use the current searching tools to search for review candidates, building our 
%let the users vote for the rules

\subsubsection{Rules for Filtering Out Less Informative Reviews} 
%Following prior work \cite{10.1145/3308558.3313532} focusing on generating product description toward general users, 
We first established several simple rules to {\it filter out} review sentences that cannot be used to further the understanding of visual attributes with specific rules: 
% Considering our analysis of linguistic differences since our goal is different with selecting reviews as part of general description, we formulated our rules for filtering as:

\one Short: sentence of 3 words or fewer; \eg ``\textit{satisfied}'', ``\textit{like the shape}'', ``\textit{poor logo}";

\two Reference to the image: reviews that comment whether the actual product is consistent with the photo provided by the seller, \eg ``\textit{looks exactly as pictured}'' ``\textit{not shown as picture}''. These sentences mentioned by our participants as ``\textit{useless}'' and ``\textit{annoying}'' since they often occur in the reviews but contain little useful information for BLV customers. 
% We set the blacklist including the similar representations.
%\xac{one thought: if the review says ``it's totally different from the image'', it might be a little more useful (at least the user can decide not to trust this product)}

\subsubsection{Rules for Basic Visual Attributes}
All reviews containing the query keywords are included as candidates and annotated using basic Part-Of-Speech tags \footnote{http://www.nltk.org/book/ch05.html}, which are also known as word classes or lexical categories, \eg noun, adjective.
Based on the aforementioned analysis, a group of three researchers iteratively established several rules for extracting informative reviews as follows. 

\textbf{Rule 1: Adjective + Keyword or Keyword + Verb + Adjective.}
The descriptive adjectives usually provide supplementary visual information. 
Examples include
%``\textit{a bright blue}'', ``\textit{the royal blue}", 
``\textit{a shimmery purple}", 
%``\textit{a glossy pink}'', ``\textit{beautiful vibrant green}''
``\textit{crescent shape}'', ``\textit{a very nice-looking etched logo}'', ``\textit{squarish shape}'', ``\textit{The bubblegum color is glossy and fun.}'', etc.
It is hard to enumerate all the possible descriptive words, instead we decided to obtain qualified snippets by differentiating those with evaluative adjectives.

The evaluative adjectives expressing subjective emotions can be vague hence not helpful to further understand the visual attributes. \eg \code{nice shape}, \code{great color}.
We collected a list of such vague adjectives, including \code{great}, \code{nice}, \textit{good},  \code{bad}, \code{horrible}, \code{disappointed} etc. These snippets are less informative but still provide subjective opinions, so we still keep them as candidates, but as the lowest priority among reviews following our rules.

\textbf{Rule 2: 1st pronoun + ... + Keyword + ... + that/which/but/because.}
Rather than the simple sentences only containing expressions on attitude \eg ``\textit{I feel disappointed at the color.}'', the sentences with clauses usually provide more detailed and useful information. Coordinating conjunction, \eg with ``\textit{but}'' emphasizes the two statements contrasting or contradicting each other. Subordinating conjunctions, \eg with ``\textit{because}'' often contain detailed reasons for customers' attitude towards the visual attribute.
Examples include ``\textit{I love the color (Bubblegum), which I bought because it was the lowest cost for a color that would be difficult to misplace or forget while traveling}''.

\textbf{Rule 3: Comparative Expressions.}
Other than the common rules above, there exist some specific rules that work for particular visual attributes.
(1) Keyword (shape) + ``like/liked'', \eg ``\textit{shaped like a Cola Bottle}''. Comparing the shape of a product with a familiar daily object can be helpful for better understanding the shape; 
(2) Keyword (size) + ``fit/for/of'', \eg ``\textit{size fits in all cup holders}'', shows reviews containing details on how the product fits in the settings are informative;
(3) ``than/more of'' + Keyword (color), \eg ``\textit{it is a terra cotta than mocha}''.
Some products look different with the picture provided by the sellers. Sighted customers complaining about this kind of difference can also be informative for better understanding the actual appearance.

%``\textit{A light aquamarine that is very mild and unoffensive}'' 

%\item {\bf Coordinating or Subordinating Conjunctions}.
%Examples include ``\textit{Main lettering on chest is printed but the sleeve logo is embroidered}'', ``\textit{I love the color (Bubblegum), which I bought because it was the lowest cost for a color that would be difficult to misplace or forget while traveling}''. Coordinating conjunction using ``\textit{but}'' means the two statements contrast or contradict each other in some way. Subordinating conjunction using ``\textit{because}'' always containing detailed reasons for customers' attitude towards the visual attribute. These sentences are more informative and can be used as references.

%\xac{moved here instead}

 \begin{figure*}[ht]
  \centering
  %\vspace{-0.6cm}
%   \setlength{\belowcaptionskip}{-0.8cm}
  \includegraphics[width=\textwidth]{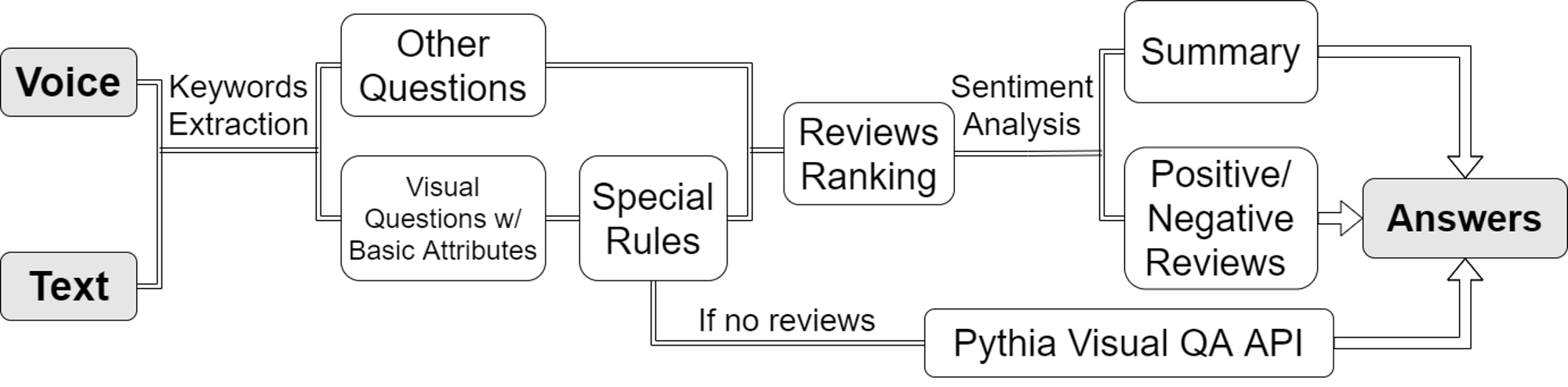}
  \caption{System overview of how Revamp responds to users' query. It first extract the keywords and process the query by its kind. After filtering out irrelevant reviews and reranking, Revamp uses the shortest candidate among the top-3 reviews to generate the visual description of a product and relevant answers to the query.
  }~\label{fig:system}
  \Description{This figure is a flowchart of the system. After keywords extraction, the users' query is divided to two groups: visual questions with basic attributes, and other questions. For the former group, we use the special rules for reviews ranking. If there is no related reviews, we leverage Pythia Visual QA API to provide the answer.}
 \end{figure*}

\subsection{Generating Answers and Image Descriptions}
Finally, we use the review snippets retrieved by the rules above to answer a BLV user's specific visual question or to compose a brief appearance description. 
As an answer to a BLV user's question, we provide original reviews retrieved by our rules divided into two lists---positive and negative---based on a sentiment analysis. We also provide a summary with the numbers of positive and negative reviews, and the top-3 informative review snippets from reviews across the two lists.
To generate both the lists and the top-3 summary, we need to rank the review snippets.
In particular, review snippets selected by Rule 1 with descriptive adjs, 2, and 3 have higher priority than those by Rule 1 withe evaluative adjs;
further, reviews of the same priority are ranked by their helpfulness (votes by other customers).
The next level lower are reviews that contain the query keywords but do not meet any rules, within which they are ranked by relevance based on the concept of graph centrality following prior work \cite{10.1145/3106426.3106444}. % an unsupervised method
For the brief appearance description, we select the shortest sentence in the top-3 candidates corresponding to each visual attribute (where each visual attribute is used as a keyword) since it is advised not to use long text for image alt-text. If users are interested in learning more details, they can move forward to ask specific questions on visual attributes.% by natural language.

\section{Implementation of Revamp---An Information Retrieval System for Online Shopping}

We present Revamp, an interactive information retrieval system to improve the shopping experience for BLV users. We implemented Revamp as a browser extension that works on Amazon. 
Revamp allows BLV users to interact with a simplified product page, access image description composed of relevant reviews, ask questions and receive a summary with original reviews as response, as shown in Figure ~\ref{fig:system}.
On the front-end, Revamp uses Chrome API to automatically simplify the page and injects our search component to the current page, allowing user to send query and get a response from our back-end.
On the back-end, Revamp extracts the keywords in user’s query and matches the keywords with review snippets data. By ranking the filtered review snippets and applying sentimental classification, the back-end returns a review summary and a positive and a negative lists.

{\bf Data source}.
Our data source consists of two main parts:
\one Basic attributes (title, color, price) from information provided by sellers;
\two Reviews (containing title, content, rating, helpfulness, date, and author); and
\three Customer Q\&A;
We scraped data from Amazon product pages using the Python library BeautifulSoup combined with Selenium and stored it in \texttt{.csv} format. 
If a product browsed by a user already exists in our database, we will use it directly; otherwise, we will run the scraper to retrieve the data and save it in the database. 
Our Python Flask based back-end API server will use these data to generate responses to our front-end's requests. Before our user study and evaluation, we pre-scraped the product data to avoid the potential latency of retrieving it in real-time.

{\bf Web page simplification}
%avoid taking the liberty and remove or replace original info from the user;
We provide an an alternate view of the original Amazon pages to improve the screen reader browsing experience. Our extension firstly rearranges the elements on the Amazon product page by manipulating the DOM tree %\xac{technically you don't need an API to manipulate the DOM tree}. 
We removed the irreverent information such as advertisements and promotions, and keep only the product details, reviews and images from the original page. Revamp would generate brief descriptions as the alt-text of the images using the aforementioned rules.
Then we use Chrome APIs to inject our Revamp module into the product page. We strictly followed WAI-ARIA\footnote{WAI-ARIA stands for Web Accessibility Initiative – Accessible Rich Internet Applications. It is a specification published by the World Wide Web Consortium (W3C) that specifies how to increase the accessibility of web pages.} standards to make sure all components have proper attributes and keyboard interactions.
%\xac{echoing Jake's comments, we need to be careful not to take the liberty and remove or replace original info from the user; rather what we provide is an alternate view of the original page}

%\subsubsection{Questions Classification}
% 
%subsubsection{Keywords Expansion}
% 
%Normally, in the information seeking process, user may try many keywords to target the information they wanted to know. e.g., If user want to know the feeling of the material of clothes, they may to try questions "How does it feel", "What is the material?", "Is it comfortable?"... This tedious process affect the shopping experience of users. The general synonyms dictionary like WordNet have a wide range of synonyms and the result searched by these synonyms are too rough, and it cannot provide the keywords pair like "feeling" and "material". To solve this, we summarized some rules for the keywords expansion and made a mini dictionary, in this dictionary, we picked the synonyms that appear frequently in the review text for each categorized questions' keywords. We already defined. e.g. Feeling has the following expanded keywords in our dictionary "material", "comfortable", "soft", "warm"... 
%待补充，如果扩充近义词，涉及到一个近义词排序的问题，很是头疼

%木有近义词

%The keywords we support for special rules: color/style/size/material %(feel)/function/quality/price
%response to special words

{\bf Review Snippets Extraction}.
Revamp responds to both keyword queries such as \code{color} as well as natural language questions such as \code{Does the product have physical buttons?}. For keyword queries on four main visual attributes, including color, logo, shape, size, we extracted the reviews with the pre-defined extended keywords based on our formulated rules. Specifically, for color, the extended keywords included `color', basic color name such as `blue' and the special name provided by sellers such as `surf the web'. 
%If the user asks about `color' without clearly mentioning the name, we provided answers based on the reviews of the color chosen by user before triggering the Revamp.
%\xac{`chosen by user before triggering the Revamp' what does this mean?}
For other natural language questions, we used the Rapid Automatic Keyword Extraction (RAKE) library \footnote{https://github.com/csurfer/rake-nltk} for keyword extraction, Synset from WordNet \footnote{https://www.nltk.org/howto/wordnet.html} to get the groupings of synonymous keywords that express the same concept, then extracted the reviews with the certain and synonymous keywords.

%Shopping websites usually ranked their reviews simply by user's votes and matching \xac{matching of what?}; however, this approach suffers from the ``Cold Start Problem'' \cite{10.1145/1352793.1352837}, as the most recent reviews will get less attention. 
%\xac{is this really an issue? i remember users can choose to see the most recent reviews}
%To make the top-ranking reviews the most relevant reviews in the reviews list, we followed MRR \cite{10.1145/3106426.3106444} method to rank the snippets by their graph centrality. 
%\xac{how does MRR work and how does it solves the recency problem?}
%By ranking the MRR score, we can get a list of candidate snippets.
%\xac{previously you also mention ranking by other customers' helpfulness rating (for reviews that meet the rules). need to be consistent.}

{\bf Answer Generation}.
%image description
%short answer
%including briefing and two review lists by sentiment
%We deployed an aspect-based sentimental classification to classify the sentiment categories of intents in reviews. 
%sentiment
%monkeylearn
There are three types of information generated.
\one A supplementary textual description of product images;
\two Two reviews lists: positive and negative; and
\three A summary of reviews based on the two reviews lists.
After we retrieved the ranked reviews, we divided them into positive and negative categories based on aspect-based sentimental classification by MonkeyLearn API.
%By default, Amazon classifies the sentiment of reviews by rating, \eg a 5-stars review will be classified as "Positive". 
%However, when a question asks for a specific aspect, \eg sleeves, a positive review might actually contain negative feedback with respect to the aspect (sleeves), \eg ``Overall is perfect... but the sleeves are not durable''.
% , and it the snippet "but the sleeves are not durable" got a controversial judgement. 
%We leveraged MonkeyLearn API to perform sentimental classification.
%By ranking the scores of intents in given keywords, we divide our review list into positive and negative sublists. 
Finally, we extract top three reviews from each list to generate our summary using a pre-defined template.
In case there are too few reviews to populate the answers, we leverage the existing computer vision technique to provide basic answers, \eg using Pythia \cite{jiang2018pythia} to answer the color, whether there is a logo on the image, and what is the shape. Users can either use text or voice input, when Revamp finishes generating answers, it will notify the user with a beep. Rather than automatically reading out the answers, we support users to use their own screen readers to scan the answers, which provides them with more freedom of control.

% figure 1
%\eg ``I found xxx reviews, xxx of them are positive, including 1..., 2... and 3..., and xxx of them are negative, ...".

%An image description gives more details than alt text and allows someone to learn more about what is in an image that goes beyond alt text. Alt text gives the user the most important information while image descriptions provide further detail. For example, alt text tells someone that there's a puddle on the floor, and image description tells someone that the puddle on the floor is in the middle of the floor and it's orange juice.

%"I’m always excited to see people include alt text and image descriptions for the visually impaired in their product descriptions. This helps me ensure that I know what I am buying, and I don’t end up buying something that isn’t my style in the slightest. By knowing how to write image descriptions about online products for visual impairment, customers can feel more confident and independent when making purchases."
\section{Evaluation}
We separated evaluation into two parts: rule evaluation and system evaluation.
Rule evaluation is to evaluate the performance of rules-based solution for retrieving informative review snippets.
System evaluation is to further evaluate how Revamp is integrated into BLV users' online shopping process. Modifying page structure and reviews-based interactions are non-separable, as they are coordinated to support a comprehensive information seeking flow. Hence we chose to evaluate Revamp as an integrated system.

\subsection{Rule Evaluation}
\begin{figure}[t]
  \centering
  \includegraphics[width=1.0\columnwidth]{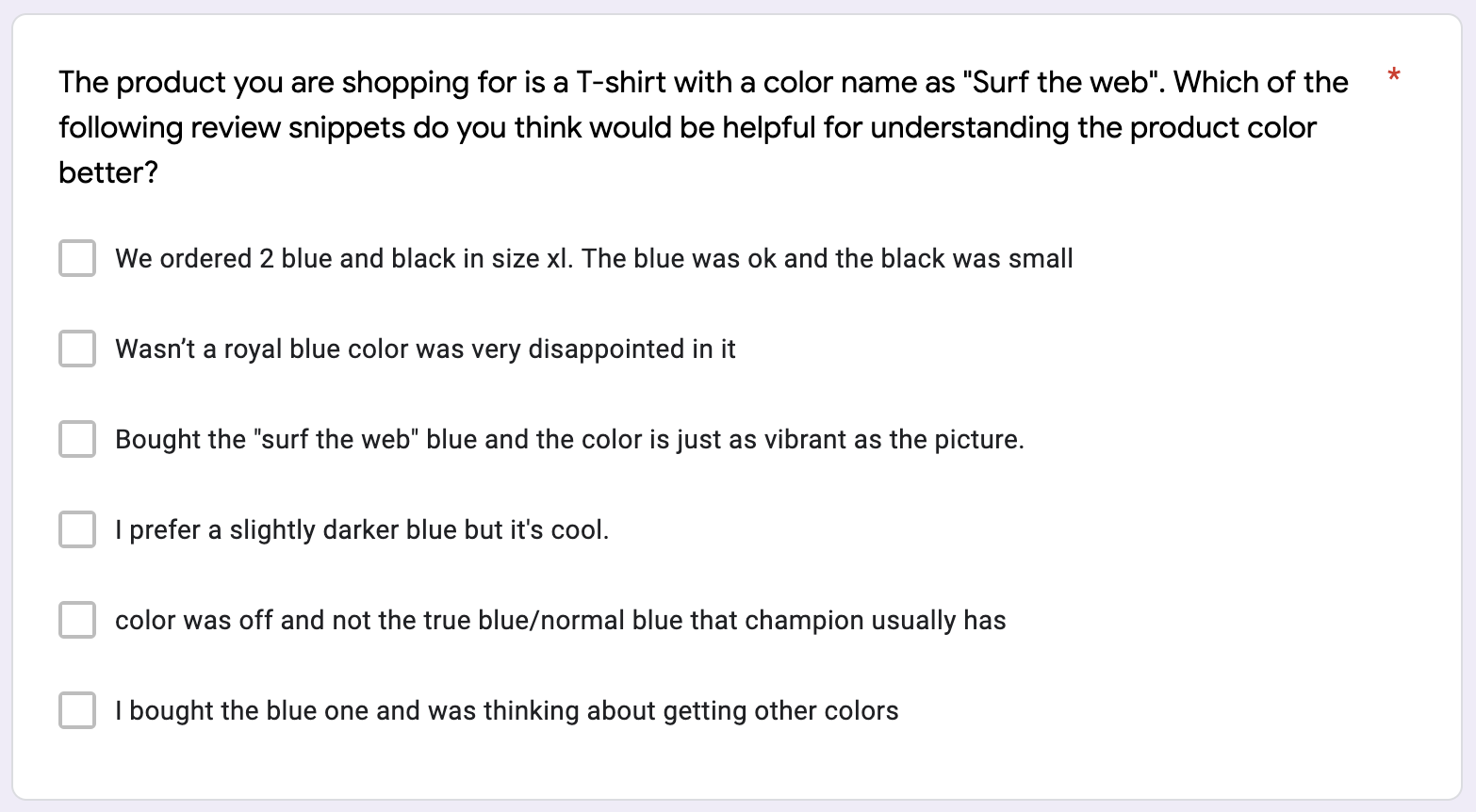}
  \caption{An example questionnaire for BLV users to vote for which one of the review snippets (retrieved by our rules or by default ranking of search result) best answers visual attributes questions. 
  % and the voting numbers of each snippet.
  }~\label{fig:vote}
  \Description{This figure is a screenshot of the questionnaire. The task is to choose which of the review snippets would be helpful for understanding the product color ``Surf the web''. Six review snippets are provided as options.}
\end{figure}

% We conducted preliminary validation on how our solution 
We first evaluated how our rules cover the informative review snippets for understanding appearance and how our solution can be generalized to different products.
% \xac{describe early on how voting worked.}
\subsubsection{Coverage of Informative Reviews for Understanding Appearances}
We randomly selected eight products covering three aforementioned main categories from the Amazon best-seller list and extracted review snippets covering four main visual attributes (color, logo, shape and size) following the above rules. We then distributed voting questionnaires to eight BLV participants (P13 - P20).
Given each product, a participants was asked to vote for the snippet that could help them best understand the visual attributes of that product:
half of the snippets were generated by our proposed rules and the other half from the top-ranked answers in the existing Amazon `Have a Question' section; all snippets were presented in a random order.
Figure ~\ref{fig:vote} shows an example question and review snippets.
% \xac{move fig closer}
The snippets selected by our rules gained 85\% of the 144 votes. Notice that we are not filtering out the remaining 15\%.
%The goal is to rank the reviews providing further insights for understanding the appearance as the top-relevance ones to improve the efficiency of information seeking.
Review snippets which were not selected by our rules may also contain informative details, but seldom directly for describing the visual attributes, \eg a review \code{We ordered 2 blue and black in size xl. The blue was ok and the black was small} talking about color actually gave more details about the size differences between different colors. Rather than filtering out these reviews, we still show them but with lower ranking, while reviews that meet our rules and provide descriptive details such as \code{color was off and not the true blue/normal blue that champion usually has} will be assigned a higher priority.
% After the voting, the researcher introduced the syntactic rules and asked if these rules were reasonable to them.
% All participants express surprise on the reasoning behind the rules, as P18 commented \code{These rules are super clever. I didn't imagine that they were effective.}
%\xac{i removed the last bit of participants' reaction becaues it is anecdotal and is likely dismissed by reviewers.}

\subsubsection{Generalization to Various Products}
\begin{figure*}[t]
  \centering
   %\vspace{-0.6cm}
   %\setlength{\belowcaptionskip}{-2.3cm}
  \includegraphics[width=2.0\columnwidth]{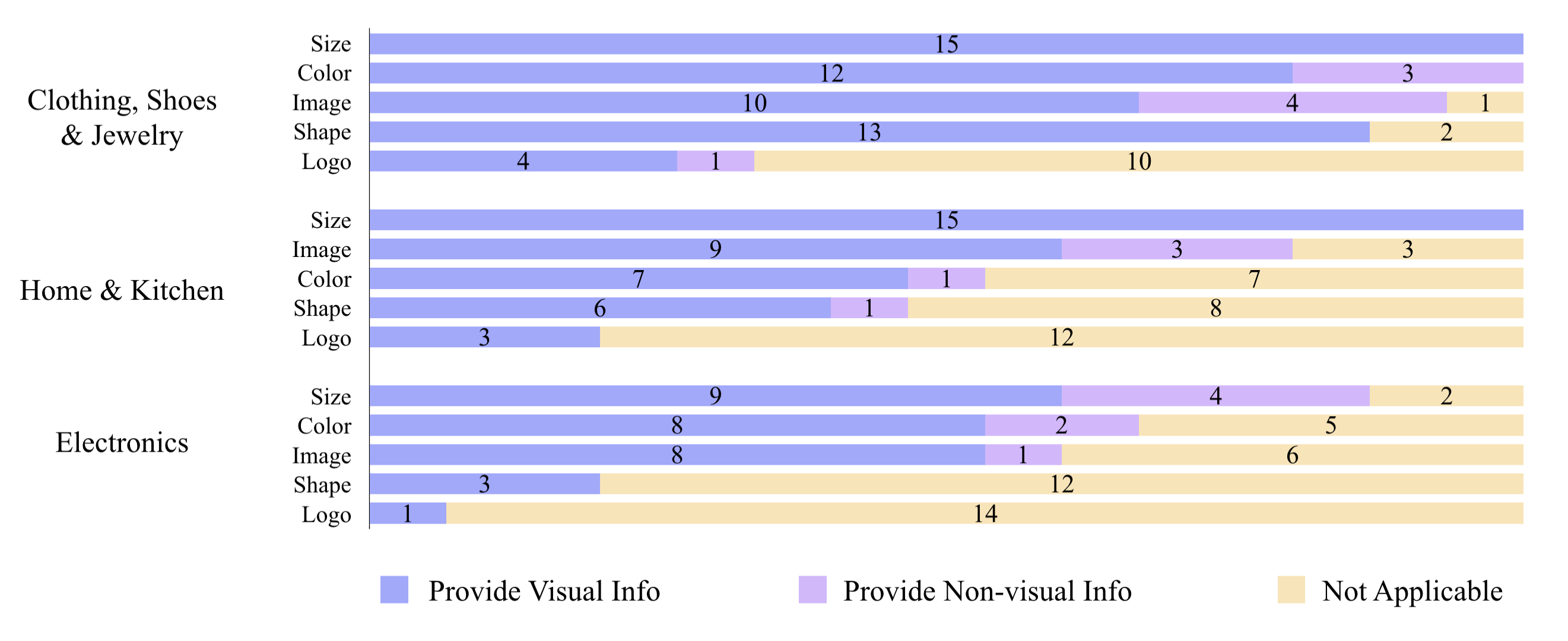}
  \caption{Performance of image description and visual questions on four basic attributes including shape, logo, size, color. Two sighted research team members rated the quality by three levels on 15 Amazon top-seller products in each of the three main categories. Results showed that our rules can directly provide visual info for 54 questions or image descriptions in Clothing, Shoes \& Jewelry; 40 in Home \& Kitchen; 29 in Electronics. We also included ``Not Applicable'' because of no relevant reviews or too few informative reviews.
  }~\label{fig:eva1}
  \Description{This figure contains histograms of performance of the generated answers on the three product categories.}
\end{figure*}

%To gain an overall understanding on how our rules can cover reviews in a wider range, we first tested our rules on the Amazon Review Dataset 2018 \cite{ni2019justifying} that includes reviews between May 1996 and Oct 2018. Results showed that our rules match 11\% reviews in Home \& Kitchen (total: 21,928,568), 15\% in Clothing, Shoes \& Jewerly (total: 32,292,099), and 9\% in Electronics (total: 20,994,353). \xac{maybe reporting the absolute numbers, \eg matching 12345 in clothing (xx\%)}.
%\xac{maybe a more interesting stat is on average, how many reviews are matched per product?}
%Although the numbers were not huge, we ensure that every review extracted by the rule was helpful \xac{how ensure? we didn't check each review here did we?} and contains the information we wanted. In other words, our rules value on precision rather than recall. \xac{we can't prove recall either because there is no groudtruth}
To further validate the generalization of these rules in a more real-world setting,
%  while under the limitations of crawler and computing resources, 
we downloaded web data of 45 selected products from Amazon's up-to-date top-seller ranking list (top 15 in each of the aforementioned three categories with similar products removed to maintain variety). 
Two sighted research team members then browsed the product page and rated the quality of the generated alt-text and answers that describe four visual (color, logo, shape, size)
% and five non-visual (quality, material, price, comfortable, button) queries 
in three levels: \textit{not applicable, providing related non-visual information, providing direct visual information}. Results showed that our rules can provide informative results in most cases as long as there exist enough reviews, especially for the products whose appearances matter more in shopping decision, as shown in Figure ~\ref{fig:eva1}.
% and \ref{fig:eva2}. 
Not applicable is due to no reviews mentioned the attribute if the specific attribute does not occur (there is no logo on the product), or has no details to provide (\eg no reviews of a white bed sheet mentioned color details), or there exist too few informative reviews. 
In this case, Revamp uses Pythia Visual Question Asking \cite{jiang2018pythia} to provide back-up answers.

% \begin{figure}[t]
%   \centering
%    \vspace{-0.6cm}
%    %\setlength{\belowcaptionskip}{-2.3cm}
%   \includegraphics[width=1\columnwidth]{eva2.png}
%   \caption{Performance of non-visual questions on four topics including Quality, Material, Price, Comfortable. For non-visual questions, we do not provide visual information from reviews, thus the results were only evaluated at two levels.
%   }~\label{fig:eva2}
% \end{figure}

\subsection{System Evaluation}
\begin{figure*}[t]
  \centering
   %\vspace{-0.6cm}
   %\setlength{\belowcaptionskip}{-2.3cm}
  \includegraphics[width=2.0\columnwidth]{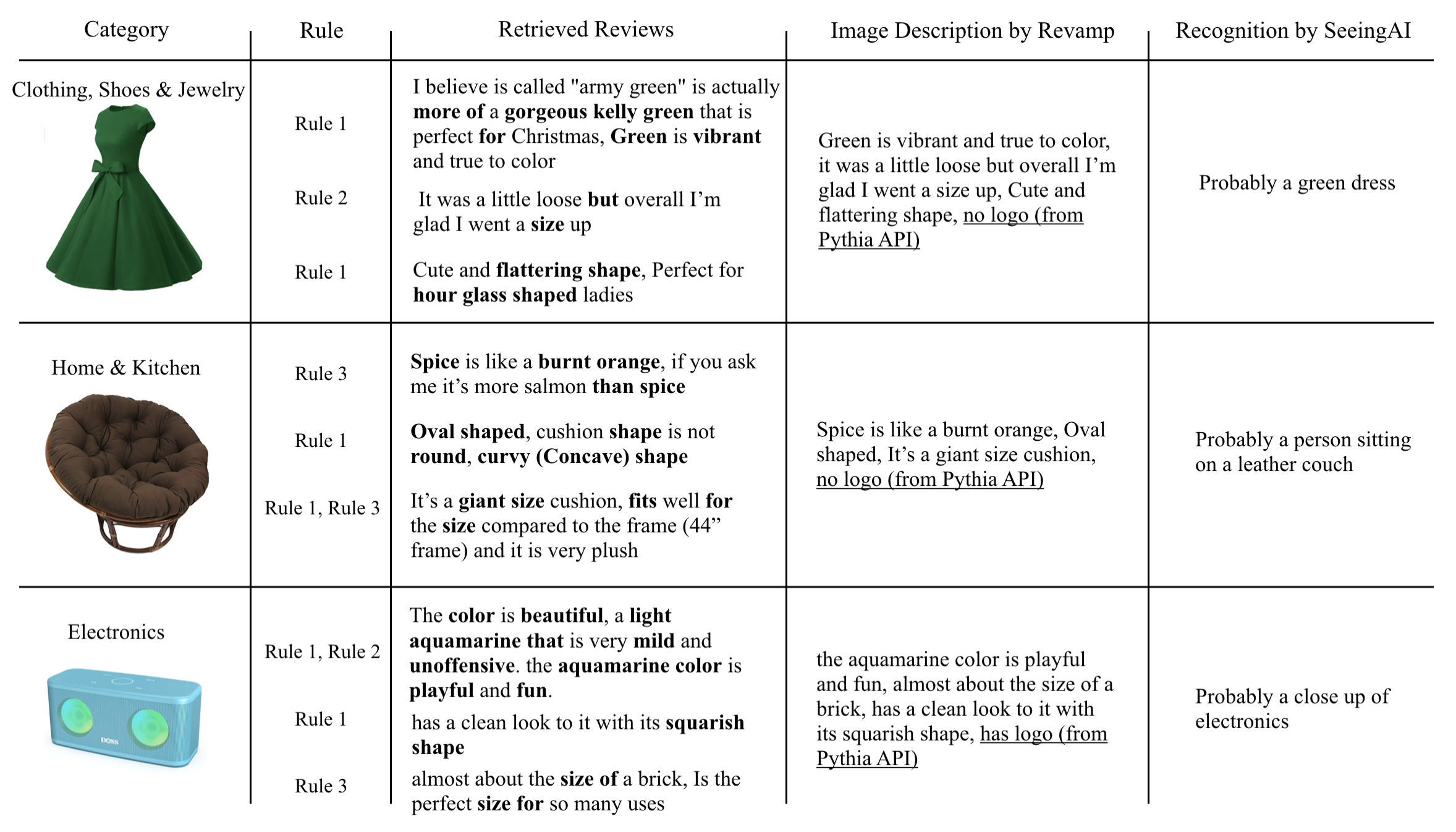}
  \caption{Examples of representative products in three main categories, showing the retrieved reviews, generated image description and comparing with the recognition results by SeeingAI, a common tool for BLV users to figure out what is in an image.
  }
  ~\label{fig:example}
  \Description{This figure shows three representative products,including a dress, a chair cushion and a speaker, and the camparison of image description generate by Revamp and Recognition result by SeeingAI.}
\end{figure*}

To further understand how the system is integrated into BLV users' online shopping process, we evaluated Revamp with BLV consumers. 
We selected six representative products from the aforementioned three main categories
%\xac{is that correct?} 
% that participants buy in daily life, 
including Water Bottle, Men’s Tee, Women's Dress, Bed Sheet, Bluetooth Speaker, and Chair Cushion as the testing products. Three representative products, retrieved reviews by rules and generated image description are shown in Figure ~\ref{fig:example}.

\subsubsection{Participants, Design, and Procedure}
We conducted interviews with the same eight blind participants (P13 - P20). Each user study lasted approximately 40 mins using Zoom after we received the IRB approval.
Participants used their own laptops and screen readers and shared their feedback with the experimenters at the same time. 
We first asked the users to browse the products as they used to on original Amazon web pages without Revamp. 
The main task was to obtain information on four visual attributes (color, logo, shape, size) also other details they were curious about.
We then introduced how Revamp worked, including how it simplified the original Amazon web page and what kinds of questions Revamp could answer, and asked the participants to repeat the same task using Revamp.
Participants then filled out a survey with Likert-scale statements as shown in Table ~\ref{tab:statements}. Participants then further explained their reasons of giving each specific ratings. Each study was audio recorded and transcribed and participants' qualitative response was summarized by affinity diagramming.

\begin{table*}[h]
\caption{We collected participants' subjective ratings on Revamp. The scale was 1 to 7, in which 7 = I strongly agree with this statement, 4 = neutral, and 1 = I strongly disagree with this statement. The value represents the average of ratings (SD). Data was analyzed using Wilcoxon test and a statistical significant difference ($p < 0.05$) is marked with $*$---all ratings of Revamp significantly outperformed the current practice.
}
~\label{tab:statements}
\Description{This table contains six statements on the user experience with or without Revamp.}
\begin{tabular}{lcc}
\hline
\multicolumn{1}{c}{\textbf{Statements}}                                                                                                                                  & \multicolumn{1}{l}{\textbf{Revamp}} & \textbf{Current} \\ \hline
It is easy and efficient for me to locate product-related information I need *                                                                                            & 5.50 (1.07)                         & 4.75 (0.89)      \\
\multicolumn{1}{l}{My questions are answered with informative answers *}                                                                                                 & 5.13 (0.64)                         & 4.38 (1.41)      \\
\multicolumn{1}{l}{I feel confident in understanding the appearance of the product *}                                                                                    & 5.88 (1.25)                         & 3.00 (0.76)      \\
\multicolumn{1}{l}{I believe I can fully utilize the information from reviews *}                                                                                         & 5.75 (1.83)                         & 5.13 (1.46)      \\
\multicolumn{1}{p{0.7\textwidth}}{Revamp can be a helpful alternative for shopping online independently when no sighted helper is available} & \multicolumn{1}{c}{6.50 (0.53)}     & -                \\
\multicolumn{1}{l}{I will use Revamp regularly in my daily life}                                                                                                         & 6.00 (0.76)                         & -               \\
\hline
\end{tabular}
\end{table*}

\subsubsection{Results}
We observed the common interaction pattern with Revamp proceeded as follows: The participants first navigated through four elements (product name, image with alt-text, info provided by sellers and QA) on Revamp. They then asked questions on visual attributes and other aspects they were curious about (no predefined questions were provided). Since personal interests differ, they asked 2 \textasciitilde 5 questions for each product, including both visual and non-visual questions. Besides the review summary, they also browsed the positive and negative review lists when needed.

\begin{figure*}[h]
  \centering
  %\vspace{-0.6cm}
  %\setlength{\belowcaptionskip}{-4cm}
  \includegraphics[width=2.0\columnwidth]{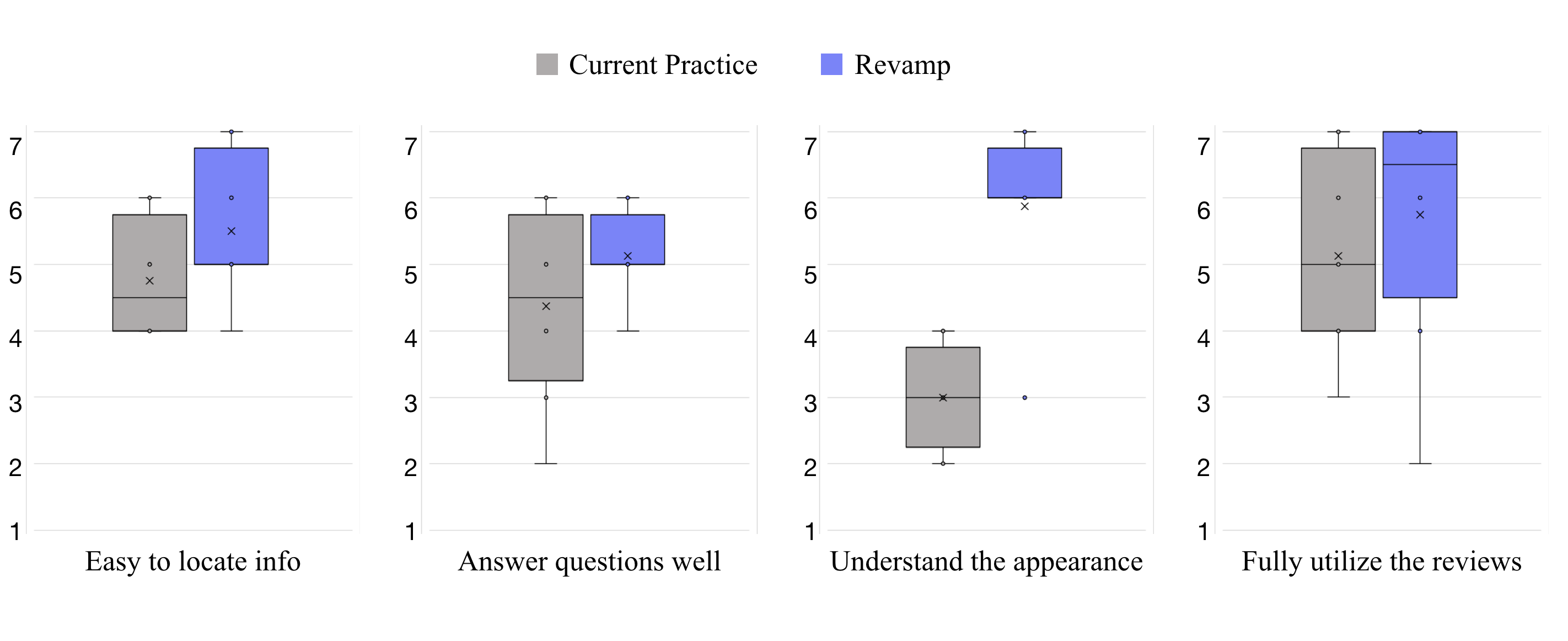}
  \caption{Subjective ratings on statements comparing current practice with experience of using Revamp. Revamp demonstrates a clear advantage in the experience of understanding the product appearance.
  }~\label{fig:evaluation}
  \Description{This figure is a violin plot to show how the user experience is improved with Revamp in four aspects: 1) Easy to locate info; 2) Answer questions well; 3) Understand the appearance; 4) Fully utilize the reviews.}
 \end{figure*}

Overall participants considered Revamp to be a helpful alternative for shopping online independently when no sighted helper is available and will use Revamp regularly in daily life. The comparative ratings are shown in Figure ~\ref{fig:evaluation}. Notice that one limitation of study is the bias of comparing a new system to an old one, participants would know which system was designed by the authors and can be influenced by this. Unfortunately, given the popularity of Amazon, participants would have known which one was ours even if we had intentionally de-identified the two systems.
We further discuss participants' subjective comments below.

{\bf Providing supplementary information for better understanding the product appearance}.
With Revamp, participants' ratings on their understanding of product appearance were improved comparing with their current practice, owing to the supplementary image description and responses to visual questions. 
Having a supplementary description of the product image is a new experience to them: \code{You know the Amazon doesn't provide image descriptions for the products. I will definitely use this add-on in my life.} (P13)
When using Amazon without Revamp, participants often just quickly skipped the image web component since no informative information is provided; with Revamp, participants tended to use the shortcut command of screen readers to directly jump to image to first gain an overall understanding of the product's appearance.
%P14 commented \code{...but sometimes I forgot to check the image if you (the researcher) didn't remind me, I just used to skip them.}

Participants felt surprised when Revamp provides useful answers on visual attributes, as mentioned by P19: \code{It is really cool. It did answer my questions about product appearance.} and even helps them learn about unfamiliar visual attributes, as shared by P13: \code{Although I haven't seen colors before, I have a lot of fun in reading these descriptions of colors. For example, I don't know what is the `Spice' color. It told me a review mentioned `like a burnt orange', which is much more understandable.}
Participants also mentioned that descriptive details of reviews from the first-hand buyers sometimes felt more trustworthy than opinions of a friend or family member, as P15 commented \code{Who can perform better than the customers who have bought the product themselves on describing the product? I really like the idea of using the reviews.} Most participants agreed that they can better utilize the information from reviews. Only one participant gave a low rating of 2 because he personally preferred not to use any filter on the reviews in case something important is missed.

{\bf Enhancing interaction flow of information seeking}.
With Revamp, the information seeking experience of participants was improved comparing with their current practice, owing to the reconstructed web page and better utilizing reviews.
Revamp provided users with cleaner web page structure and is more user-friendly than the current Amazon page, as stated by P20: \code{I don't need to worry of being stuck in useless information any more.}
Participants found the review summary and the review lists divided by positive and negative useful to access customers' opinions more efficiently and proactively. P14 mentioned that ``\textit{I like it that Revamp also keeps the original reviews accessible in the lists. After reading the summary, I can then make the decision to skip or look into the details of each relevant reviews in the list.}'' Participants were more engaged in asking questions, as P19 commented \code{Sometimes I could be inspired to ask more after I read the answers.}
Participants could choose to interact with the system using either voice commands or text input based on their preferences. 
% No matter which modality they chose, participants liked the experience of screen readers augmented with generated answers. 
Although participants in general liked the experience of screen readers augmented with voice input, they preferred to receive answers in text rather than speech:
\code{Using my screen reader to read out the answers is far more better than directly answering my questions by voice. I can adjust the speed and pause anytime.} (P15)

\section{discussion}
Most web pages with vivid designs and a large amount of information are still not accessible enough to BLV users at present. We explored three aspects towards enhancing the information seeking experience on online shopping websites: \one{simplifying and reconstructing the web pages according to users' current task;}  
\two{providing coordinated experience of active query and passive reading to support flexible information seeking;}
\three{leveraging relative text resources on the web page, such as reviews, to fill in the information gap.} Besides, this work also inspired several exciting future directions as follows.

\subsection{Working with Multiple Product Pages of Online Shopping}
In this work, we focus on the task of understanding the product details, especially the appearance. Imagine another task: if the user has several product candidates in mind, we should also correspondingly meet a new information seeking need of comparing multiple products. To address this, future work can explore reconstructing the webpages to support efficient switching among products and answering comparative questions about multiple products. Currently, table is a common web page element for comparing different products and should be supported in future versions of Revamp. For example, the system should guide the user to navigate a table by informing them which products are being compared in terms what attributes and further be able to answer a user's question by looking up the table.

\subsection{Working with A Broader Variety of Product Pages}
Furthermore, future work can employ and test our methods on other product pages where there is often an overload of information inaccessible for BLV users to retrieve. For example, on Yelp, our methods can add description to popular dishes, whose images do not always convey all important information (\eg how large is the portion); on Trip advisor, images taken by fellow travellers can be further described using our methods by extracting relevant descriptions from others' comments (\eg whether a trail has shade); on Youtube, comments can be leveraged to generated video captions for eye-catching moments, which serve as a more vivid introduction before one decides whether to watch a given video.

\subsection{Working with Supervised methods}
%Another challenging research question touched by this work is how to fill in information conveyed by image to BLV people using text descriptions.
Although the subjective comments were useful and convey vivid details, our participants pointed out that the generated appearance descriptions might be too subjective since it only contained one snippet in each visual aspect. 
However, if we follow prior work on filtering out subjective comments \cite{10.1145/3308558.3313532}, we will lose the vivid and descriptive details that are crucial to appearance understanding.
As such, participants preferred the image description to remain as concise as possible and the description provided by Revamp would serve more as \code{a first step or a clue.} (P16) 
Also, it remains to be explored how we can better retrieve useful information from reviews on the high-level visual concepts inferred by image. 
In this work, we formulated hand-crafted rules to explore the possibility of leveraging reviews as an additional information source to fill the visual information gap.
In the future, we can collect and label data to deploy supervised methods such as Knowledge Graph which could extract more information to better support queries not limited in the four visual aspects (color, logo, shape and size) and provide better summative appearance description.

\subsection{Working with Better QA Experience}
The question \& answering experience of Revamp can be improved by providing prompting questions and involving more information sources. We observed that some participants had difficulty in formulating what questions to ask: P16 only asked about basic attributes, \eg color, quality. She explained that \code{Sometimes I don’t know how to ask a `good' question. Maybe if I change my wording, I can get more answers.} and suggested us to provide more pre-defined questions as prompts. Meanwhile, participants noticed that there still exists a gap between their expectations and the answer quality in other details beyond the four main visual attributes. ``\textit{Revamp can support descriptive answers for basic visual questions, but when I want to ask about some concrete (non-visual) details such as the dimensions of the bottle, it cannot answer me directly.}''(P19)  In the future, we can involve more information source into Revamp, also leveraging Optical Character Recognition to extract the text information on the images provided by sellers.

\subsection{Working in the Real-World Settings}

Similar to other information retrieval system, the performance of Revamp highly depends on the quality of existing data. Currently, we leverage existing visual question answering API to provide a back-up answer for visual questions when there are too few reviews or no informative reviews. In the future, we can also involve human helpers in the system by sending the questions to crowd workers.
Our rules for extracting informative reviews can also give insights on formulating guidelines of describing products for BLV users: \one{include descriptive details rather than vague opinions (Rule 1);} \two{if have to express personal attitudes, elaborate the reasons (Rule 2);} \three{compare with daily or common objects to help understand new concepts (Rule 3).}
Besides, participants also hoped to see Revamp's functions integrated into the mobile application with Alexa since they often try the mobile application of Amazon when faced with accessibility problems shopping on the web.

\subsection{Limitations}

There are several limitations of the system evaluation study. First, we did not counterbalance the conditions in the system evaluation. Using Revamp first would introduce a strong carry-over effect, as the information provided by Revamp is an augmentation of the baseline. As personal interests and product information differ, to compare intuitively how Revamp improves the experience, all participants first browsed the products as they used to on original Amazon web pages then on the modified web page of the same product by Revamp.
Second, there exists the bias of comparing a new system to an old one when participants know which system was designed by the researchers \cite{10.1145/2700648.2809867, 10.1145/3313831.3376749}. Third, only eight participants and a limited amount of products were involved in evaluation. This limitation can be addressed by a large-scale in-the-wild study, which we regard as future work beyond this initial study.

%\xac{other solutions if there are too few reviews or no informative reviews}

%\xac{future work on extracting functionality-related information from reviews}

%\xac{look back at CHI 2020 and UIST 2020's reviews---there might something we can discuss about}

%\xac{discuss how to address participants' concerns in future work}
\section{Conclusion}
We present Revamp, a system that employs information retrieval techniques to meet the unique information seeking requirements of BLV consumers when shopping online. 
% Revamp provides a coordinated screen reader and question-answering experience. 
Our main contribution is a rule-based approach that leverages rich customer reviews to serve as image description and to answer BLV users' questions related to product appearances.
Evaluations with eight BLV consumers showed that Revamp provides useful subjective information for understanding the product appearance and enhances the accessible information seeking experience of online shopping. Although Revamp could not provide answers for all of the products (\eg when there are too few reviews of a product), it can serve as an effective supplemental helper for BLV users to better access and understand a product before making a purchase decision.

\begin{acks}
The authors thank all participants who generously shared their time and experience for this work, Mengqi Li for helping with formative studies, and Prof. Jacob O. Wobbrock for valuable advice on the paper writing.

\end{acks}

\bibliographystyle{ACM-Reference-Format}
\bibliography{sample-base}

% \balance
\newpage
\onecolumn
\newpage
\appendix
\section{Demographics of study participants}
\label{appendix:a}
\begin{table*}[h]
\vspace{1em}
   \begin{tabular}{ccccc}
     \toprule
     ID & Age/Gender & Visually impairment & Device\\
     \midrule
     \ P1 & 24/F & Blind (total), acquired & smartphone\\
     \ P2 & 24/F & Blind (some light perception), born & smartphone\\
     \ P3 & 22/F & Blind (some light perception), born & smartphone, laptop\\
     \ P4 & 25/M & Blind (total), born & smartphone\\
     \ P5 & 24/M & Blind (some light perception), born & smartphone, laptop\\
     \ P6 & 24/M & Low Vision (narrowed field), born & smartphone, laptop\\
     \ P7 & 27/M & Low Vision (narrowed field), born & smartphone\\
     \ P8 & 32/M & Blind (total), born & smartphone, laptop\\
     \ P9 & 28/M & Blind (total), born & smartphone, laptop\\
     \ P10 & 22/M & Blind (total), acquired & smartphone, laptop\\
     \ P11 & 27/M & Blind (total), born & smartphone,laptop\\
     \ P12 & 23/F & Low Vision (narrowed field), acquired & smartphone, laptop\\
     \ P13 & 34/M & Blind (total), born & smartphone, laptop\\
     \ P14 & 35/M & Blind (some light perception), acquired & laptop\\
     \ P15 & 33/M & Blind (total), born & smartphone, laptop\\
     \ P16 & 22/F & Blind (some light perception), born & smartphone, laptop\\
     \ P17 & 27/M & Blind (some light perception), born & smartphone, laptop\\
     \ P18 & 31/M & Blind (total), born & smartphone, laptop\\
     \ P19 & 26/F & Blind (some light perception), born & smartphone, laptop\\
     \ P20 & 44/F & Blind (total), acquired & smartphone, laptop\\ \hline
      \multicolumn{3}{l}{\textit{P1 - P10: Chinese Users P11 - P20: North America Users}}
    \end{tabular}
\end{table*}

\end{document}